\documentclass{aa}
\usepackage{graphicx}
\begin{document}
   \title{Pristine CNO Abundances from Magellanic Cloud B Stars}

   \subtitle{I. The LMC Cluster NGC\,2004 with UVES\thanks{Based on observations carried out at the European Southern Observatory (ESO), Paranal, Chile under programme ID 66.D-0214(A)}}

   \author{A.J. Korn\inst{1}
          \and
	  S.C. Keller\inst{2}
          \and
          A. Kaufer\inst{3}
	  \and
	  N. Langer\inst{4}
	  \and
	  N. Przybilla\inst{1}
	  \and
   	  O. Stahl\inst{5}
	  \and 
	  B. Wolf\,\inst{5}  }
   \offprints{A.J. Korn}

   \institute{Univerit\"ats-Sternwarte M\"unchen (USM), Scheinerstr.~1, 81679 M\"unchen, Germany\\
\email{ajkorn@usm.uni-muenchen.de, nob@usm.uni-muenchen.de} 
         \and
	Lawrence Livermore National Laboratory, 7000 East Ave., Livermore, CA 94550, U.S.A.\\ \email{skeller@igpp.ucllnl.org}
	 \and
European Southern Observatory, Alonso de Cordova 3107,
Santiago, Chile \ \ \email{akaufer@eso.org}
	 \and
 	Astronomical Institute, Utrecht University, NL-3584 CC, Utrecht, The Netherlands\\
\email{N.Langer@astro.uu.nl}
	\and
	Landessternwarte Heidelberg (LSW), K\"onigstuhl, 69115 Heidelberg, Germany\\
\email{ostahl@lsw.uni-heidelberg.de, bwolf@lsw.uni-heidelberg.de}
             }

   \date{Received 2001; accepted 2001}

   \abstract{
We present chemical abundances for four main sequence B stars in the young cluster NGC\,2004 in the Large Magellanic Cloud (LMC). Apart from H\,{\sc ii} regions, unevolved OB-type stars are currently the only accessible source of present-day CNO abundances for the MCs not altered by stellar evolution. Using UVES on the VLT, we obtained spectra of sufficient resolution ($R$\,=\,20\,000) and signal-to-noise (S/N\,$\geq$\,100) to derive abundances for a variety of elements (He, C, N, O, Mg and Si) with NLTE line formation.\\
This study doubles the number of main sequence B stars in the LMC with detailed chemical abundances. More importantly and in contrast to previous studies, we find no CNO abundance anomalies brought on by e.g. binary interaction or rotational mixing. Thus, this is the first time that abundances from H\,{\sc ii} regions in the LMC can sensibly be cross-checked against those from B stars by excluding evolutionary effects. We confirm the H\,{\sc ii}-region CNO abundances to within the errors, in particular the extraordinarily low nitrogen abundance of $\varepsilon$(N)\,$\simeq$\,7.0. Taken at face value, the nebular carbon abundance is 0.16\,dex below the B-star value which could be interpreted in terms of interstellar dust depletion. Oxygen abundances from the two sources agree to within 0.03\,dex.\\
In comparison with the Galactic thin disk at MC metallicities, the Magellanic Clouds are clearly nitrogen-poor environments.    
   }

   \maketitle
%

\section{Introduction}
Young populous clusters like NGC\,2004 in the Large Magellanic Cloud (LMC) are ideal objects to study both the current state of chemical evolution of the LMC and the physics of hot stars at a metallicity a factor of two below that of the Sun.\\
The study of stellar absorption spectra yields chemical abundances with an accuracy comparable to that achieved from the analysis of nebulae such as H\,{\sc ii} regions. However, both sources are subject to physical processes which alter the abundances of certain chemical species in a systematic fashion. In most cases, the {\em direction} of these systematic effects is known, yet their extent is often uncertain. In particular, the scaling of these effects with metallicity is poorly known. In H\,{\sc ii} regions, it is mostly dust formation and depletion onto grains which will result in lower limits on the present-day ({\em pristine}) abundances. As for B stars, rapid rotation can lead to contamination of the atmospheres with CN-cycled material from the core resulting in He and N enrichment and a corresponding C (and -- to a lesser extent -- O) depletion (Fliegner et al. \cite{fliegner}, Meynet \& Maeder \cite{maeder}, Heger \& Langer \cite{heger}). The onset of ``rotational mixing'' is a function of the rotation rate, stellar mass and metallicity. Rotating models of hot stars indicate that it can take place very early on, i.e. whilst on the main sequence (MS). Furthermore, one can expect rotational mixing to be more efficient in metal-poor environments (cf. Maeder \& Meynet \cite{mandm}), as metal-deficient stars have smaller radii for a given mass and thus rotate faster for a given angular momentum.
\begin{table}
      \caption[]{Chemical abundances of bona-fide MS B stars in the LMC in comparison with H\,{\sc ii}-region data. $\varepsilon$(X)\,:=\,$\log$($n_{\rm X}/n_{\rm H}$)+12. Typical errors are 0.2\,--\,0.3\,dex.}
      \label{prevB}
      \begin{tabular}{lllll}
      \hline
       star & PS 34-16$^{\mathrm{a}}$ & LH 104-24$^{\mathrm{a}}$ & NGC\,1818/D1$^{\mathrm{b}}$ & H\,{\sc ii}$^{\mathrm{c}}$ \\
      \hline
       $\varepsilon$(C) & \ \ \ 7.10 & \ \ \ 7.50 & \ \ \ \ 7.83 & 7.90 \\
       $\varepsilon$(N) & \ \ \ 7.50 & \ \ \ 7.70 & \ \ \ \ 7.59$^{\mathrm{d}}$ & 6.90\\
       $\varepsilon$(O) & \ \ \ 8.40 & \ \ \ 8.50 & \ \ \ \ 8.46 & 8.40 \\
       $\varepsilon$(Mg) & \ \ \ 7.00 & \ \ \ 7.40 & \ \ \ \ 7.35 & \ \ -- \\
       $\varepsilon$(Si) & \ \ \ 7.00 & \ \ \ 7.40 & \ \ \ \ 7.10 & 6.70 \\
       $\varepsilon$(Fe) & \ \ \ 7.20 & \ \ \ \ \ -- & \ \ \ \ 7.34 & \ \ -- \\
      \hline
      \end{tabular} 
      \begin{list}{}{}
      \item[$^{\mathrm{a}}$] from Rolleston et al. (\cite{rolleston}), $^{\mathrm{b}}$ from Korn et al. (\cite{korn}),
      \item[$^{\mathrm{c}}$] from Garnett (\cite{garnett}), $^{\mathrm{d}}$ revised in this publication\\[-6ex] 
      \end{list}
\end{table}

\noindent
Among a sample of B stars only the fastest rotators ($v_{\rm rot}>$\,200 km/s) are expected to display contaminated CNO abundances (Heger \& Langer \cite{heger}, Meynet \& Maeder \cite{maeder}). If one is interested in the pristine present-day abundance pattern of these elements, it is therefore sensible to look for and analyse the least evolved objects accessible.\\ 
So far, the faintness of unevolved MC B stars ($m_{\rm V}>15^{\rm m}$) has prevented an extensive confrontation of the H\,{\sc ii}-region CNO data with that from B dwarfs. Four stars have been studied in detail, three by Rolleston et al. (\cite{rolleston}) and one by Korn et al. (\cite{korn}, hereafter Paper I). Of the three stars of Rolleston et al. (\cite{rolleston}) one (PS 34--144) is classified as a He-weak star which we will therefore disregard in what follows. 
The CNO abundances of the remaining three MS stars are confronted with the most recent measurements from H\,{\sc ii} regions in Table \ref{prevB}: While the consistency among determinations of O abundances is very high, there is a significant positive offset ($\simeq$ 0.7 dex) of B-star N abundances with respect to the H\,{\sc ii}-region value.\\  
It is a priori unclear which of the two data sets reflects the pristine LMC N abundance. There are many possible explanations for the peculiar behaviour of N: either the H\,{\sc ii}-region value is systematically too low (note that it is --~like all the other H\,{\sc ii}-region data~-- exclusively based on 30 Dor and N44C, cf. Garnett \cite{garnett}), the B-star value systematically too high or we see the imprint of a physical process which alters one of the two data sets. Since dust depletion of nitrogen seems hard to envision (Mathis \cite{mathis}), it looks as if the B stars are to blame. And indeed, it might be that -- due to the limitations imposed on us by the use of 4m-class telescopes up to 1999 -- choosing the brightest targets compatible with MS colours resulted in preselecting stars that are preferentially rapid rotators as rapidly rotating stars will evolve to higher luminosities (Fliegner et al. \cite{fliegner}).\\ 
Whatever the scenario, we regard it as worthwhile to push the limit further towards the zero-age MS to see whether the N abundances already found are representative of LMC B stars in general.\\
In contrast to the paucity of data on dwarfs, many MC supergiants -- from spectral type O to K -- have been studied in detail (e.g. Haser et al \cite{haser}: O-type, Paper I: B-type, Venn \cite{venn2}: A-type, Andrievsky et al. \cite{andrievsky}: F-type, Hill \& Spite \cite{hill}: K-type). Obviously, in trying to understand the CNO pattern of these evolved objects the problem lies in separating all contributing signatures: abundance changes due to dredge-up episodes in the red-giant phase and potentially rotational mixing in earlier phases when the rotation rate was still high. This is particularly true for stars located in the so-called Blue Hertzsprung Gap (between the MS and the region of blue-loop excursions, cf.\ Fitzpatrick \& Garmany \cite{fitzpatrick}), whose evolutionary status is still highly uncertain.\\ 
Like in the MS B stars already discussed, the most significant signature is that of nitrogen:
Among SMC A-type supergiants Venn (\cite{venn2}) found a large scatter in NLTE (non local thermodynamic equilibrium) N abundances (6.9\,$\leq$\,$\varepsilon$(N)\,$\leq$\,7.7) which is hard to explain in the framework of the first dredge-up alone and argues in favour of rotational mixing in earlier evolutionary phases as the most likely cause. As far as abundances from other elements with less pronounced signatures are concerned, one has to worry about how severely they might be affected by systematic effects when comparing different spectral types: not in all cases the assumptions made (plane-parallel geo\-metry, stationarity, LTE) are as justified as in the case of dwarfs. For example, the carbon abundance from K supergiants seems to be systematically higher than in the hotter stars. On the whole, the general picture drawn above is, however, supported by all spectral types.\\ 
All in all, rotational mixing could help to resolve the discrepancies outlined above: adding rapid rotation, a fraction of any young stellar population will display modified CNO abundances, widen the MS, populate regions of the HRD which are traversed quickly by non-rotating stars and lead to enhanced scatter in N after the first dredge-up by mimicking a distribution of initial abundances.\\  
\begin{table}
\label{log}
\caption{Observation log for the programme stars. Nomenclature and visual magnitudes from Robertson (\protect\cite{robertson}). $t_{\rm exp}$ denotes the expsoure time, m/y the month and year of the observation. The line-free wavelength ranges 4202\,--\,4210, 4402\,--\,4409, 4493\,--\,4500 and 4680\,--\,4695\,\AA\ were used to
estimate the S/N ratios (1$\sigma$) of the rebinned spectra. Appended to this table are the two luminosity class {\sc iii} stars in NGC\,2004 from Paper I observed with CASPEC.}
\begin{tabular}{lcrrrr}
\hline
object in & m$_V$ & $t_{\rm \,exp}\ $ & S/N & m/y & $v \sin i$ \\
NGC\,2004 & [mag] & [h]               &     &     & [km/s] \\ 
\hline
B18 & 14.8 & 1.0 + 1.0 & 140 & 12/00 & 130 \\
C8 & 14.9 & 1.0 & 110 & 11/00 & 100 \\
C9 & 15.8 & 1.4 + 1.4 & 100 & 11/00 & 60 \\
C16 & 14.7 & 1.0 & 130 & 11/00 & 60 \\
D3 & 15.8 & 1.4 + 1.4 & 120 & 12/00 & 70 \\
D15 & 15.1 & 1.5 & 120 & 12/00 & 45 \\
\hline
B15 & 14.2 & 3.0 + 3.0 & 70 & 12/89 & 25 \\
B30 & 13.8 & 3 x 2.0 & 150 & 11/87 & 30 \\
\hline
 \end{tabular}
\end{table}
\section{Preselection of suitable candidates}
Since B stars are frequently fast rotators and/or show emission in
their Balmer lines (Be stars, cf.~Jaschek et al.~\cite{jaschek}), a
great effort was made to select the small number of stars which are
suitable for an abundance fine analysis: non-Be and with projected
rotational velocities $v \sin i$ below 100\,km/s.\\
Preselection of slow rotators in NGC 2004 was made from spectra obtained with the Siding Springs Observatory (SSO) 2.3m telescope using the Double-beam Spectrograph. Aperture plates were fabricated to obtain medium resolution (0.3\,\AA/pixel) spectra (4100\,\AA\,$<$\,$\lambda$\,$<$\,5100\,\AA) for 34 MS B stars.\\
To determine $v\, \sin i$ we have used a $\chi^{2}$ minimization technique which locates an optimal match between the observed spectrum and rotationally broadened synthetic spectra interpolated from a grid of models which spans 4000\,K\,$\le$$\,T_{\rm eff}\,$$\le$\,50\,000\,K and 5.0\,$\ge$ log\,$g$ $\ge$\,0.0 (Gonz\'{a}les Delgado \& Leitherer \cite{gonzalez}). More details can be found in Keller et al.\ (\cite{keller}). Principle limitations on this technique are given by the S/N and resolution obtainable with the instrument used.\\
As can be seen from Table \ref{log}, our preselection was very successful: four of the six targets (nearly 70\,\%) rotate at a rate low enough to analyse their metal content spectroscopically.

\section{UVES observations}
\begin{figure}
   \centering
   \includegraphics[angle=-90,width=8.8cm]{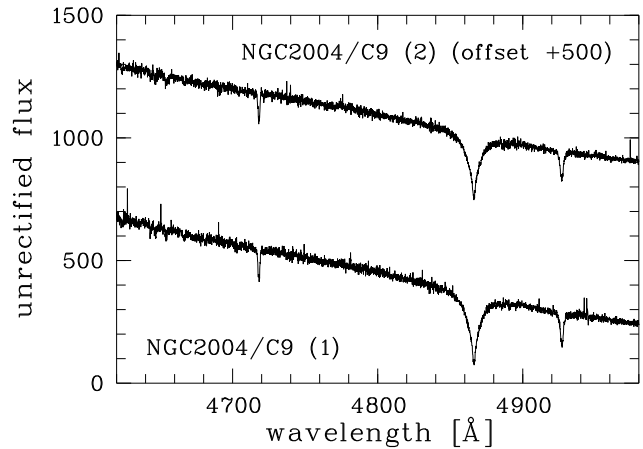}
   \caption{The red part of the Blue437 exposures of NGC2004/C9. The blaze residuals are very constant from order to order allowing for a thorough rectification using a low-order polynomial.}
              \label{continuum}%
    \end{figure}

Due to its high efficiency UVES (Ultraviolet-Visual \'{E}chelle Spectrograph) on ESO-VLT (Very Large Telescope) UT2 (Kueyen) is the ideal instrument for studying MC MS B stars ($m_V$\,$>$\,15$^{\rm m}$).
We initially considered using one of the standard templates for full optical coverage (DIC1 390\,$+$\,564\,nm), but rejected this option as it would have resulted in the loss of the strategic Si\,{\sc iii} triplet near 4560\,\AA \ (dichroic gap). We finally chose the UVES standard setting Blue\,437nm (3760\,\AA \,$<$\,$\lambda$\,$<$\,4980\,\AA ) covering all the strategic lines for a classical abundance fine analysis of B stars. The atmospheric dispersion corrector (ADC) was used and the slit oriented in such a way as to minimize contributions from neighbouring stars. The slit width was set to 1.2\,\arcsec\ (corresponding to a resolving power of $R$\,=\,40\,000) not to lose light under moderate seeing conditions.\\
One to two spectra of each of the six preselected objects were exposed in Service Mode between November and December 2000. We make use of the pipeline-reduced spectra: they turn out to be of sufficient quality in terms of rectification, even for the use of H$\beta$ as a gravity indicator. We note, though, that the blue part of the exposure (3800\,--\,4600 \AA ) is much less well-defined in terms of the continuum. After rebinning to a 2-pixel resolution of $R$\,=\,20\,000 the spectra were normalized interactively and -- in the case of two exposures -- coadded. Table \ref{log} gives a brief observation log including measured signal-to-noise ratios (S/N) and projected rotational velocities ($v$\,sin\,$i$). Both sets of quantities fall in the range expected from the UVES exposure time calculator on the one hand and the preselection on the other. Figure \ref{continuum} shows the pipeline reduction result for the two spectra of C9 in the region surrounding H$\beta$.\\
Except for B18, the heliocentric radial velocities $v_{\rm rad}^{\rm hel}$ are fully compatible with that deduced for the cluster by Freeman et al. (\cite{freeman}). B18 (just like B15 from Paper I) has a radial velocity some 40 km/s higher than the cluster average. These two stars do probably not belong to the cluster. Unfortunately, the cluster membership of the other programme stars is not certain either, since the difference in radial velocity between the cluster and the surrounding field is not significant within the errors.
\section{Analysis}
We utilize the three-step analysis developed by and extensively discussed in Gummersbach et al. (\cite{gummersbach}) and Paper~I. In short, log $g$ is derived from the profile of H$\beta$, $T_{\rm eff}$ by balancing two adjacent ionization stages of silicon as a function of $\varepsilon$(Si) and $\xi$, with $\xi$ being concordantly determined between silicon and a set of oxygen lines. In all three steps, the metallicity of the underlying atmosphere (and with it the strength of the background opacities for the NLTE calculation) is varied in a self-consistent manner. This was shown to have a non-negligible effect on the abundances to be derived. In the simultaneous determination of $T_{\rm eff}$, log $g$, $\xi$, $\varepsilon$(Si) and $\varepsilon$(O) the average underabundance of Si and O is taken to be indicative of the overall metallicity [$m$/H]. Due to the [$\alpha$/Fe] ratio inherent to the MCs (see Sect.~\ref{alpha}), this procedure results in slightly underestimated atmospheric metallicities. The residual discrepancy between [$m$/H] and [Fe/H] is, however, only of the order of 0.1\,dex. Thus this approach is clearly superior to assuming [$m$/Fe] to be solar, as it is still often found in the literature.\\
With regard to the NLTE line formation, new input physics was implemented in the model atoms for C, N and Mg. Here we only give a brief account of the expected reliability and refer the reader to the publications of Przybilla \& Butler (2001) and Przybilla et al. (2001a,b).\\
We refer the reader to Paper I for an error analysis and references to the original publication of the other model atoms. 
\subsection{Carbon}
With equivalent widths of 160\,$\pm$\,10 m\AA \ the strong C\,{\sc ii} feature at 4267\,\AA\ (multiplet 6, cf. Fig.~\ref{fits}) is clearly visible in the spectra of our MS B stars (we don't have access to the two strong lines (multiplet 2) in the blue wing of H$\alpha$). While it became clear relatively early on (Hardorp \& Scholz \cite{hardorp}) that this line yields artificially low carbon abundances in LTE, a reliable NLTE modelling was only achieved in the 1990s (Sigut \cite{sigut}). Our model atom closely resembles that of Sigut and we follow Sigut's argumentation showing that the abundances derived with it are reliable to typically 0.2\,dex.   
\subsection{Nitrogen}
The model atom has been extended to be applicable to early B-type stars. It includes the lowest 3 levels of N\,{\sc i}, 77 of N\,{\sc ii}, the lowest 11 levels of N\,{\sc iii} and the ground state of N\,{\sc iv}. Photoionization cross-sections and oscillatior strengths in N\,{\sc iii} are adopted from the Opacity Project (Fernley, Hibbert, Kingston \& Seaton; available only from the {\sc Topbase} database at http://cdsweb.u-strasbg.fr/topbase.html) and electron collisons from R-matrix calculations by Stafford et al. (\cite{stafford}).\\
The spectrum of nitrogen lines is far less pronounced than that of carbon. In fact, the strongest line at 3995\,\AA \ has equivalent widths below 50 m\AA \ in our MS programme stars (cf.\ Fig.~\ref{fits}). In combination with projected rotation rates up to 70 km/s a clear detection requires S/N ratios of around 100. The situation is less critical in stars which are more evolved and/or whose nitrogen abundance is higher, e.g. NGC 2004/B30 (125 m\AA ) or NGC 1818/D1 (74 m\AA, cf. Fig. \ref{D1}).\\
As far as the modelling is concerned, we note that the subordinate lines at 4621 and 4630\,\AA\ yield abundances fully compatible with those derived from N\,{\sc ii} 3995. As in the case of carbon, these abundances are expected to be accurate to 0.2\,dex.
\subsection{Magnesium}
The only Mg line we have access to in the B-star temperature regime is the Mg\,{\sc ii} doublet at 4481\,\AA\ (cf. Fig.~\ref{fits}). Departures from LTE are found to be $\simeq-$0.1\,dex, the absolute abundances have not changed in going from the old to the new model atom.  

   \begin{table*}
      \caption[]{Derived stellar parameters and abundances for the
      four programme stars plus NGC\,1818/D1 from Paper I. Solar System composition from Grevesse \& Sauval (\protect\cite{grevesse}), LMC H\,{\sc ii}-region abundances from Garnett (\protect\cite{garnett}). The customary notation $\varepsilon$(X)\,:=\,log($n_{\rm X}/n_{\rm H}$)\,$+$\,12 is employed. Representative error estimates are assigned to the individual abundances determinations.}
         \label{results}
\renewcommand{\arraystretch}{1.1}  
      {\large
      \begin{tabular}{lrrccrrrrccc}
\hline
object & $T_{\rm eff}$ & \ $\log g$ & $\xi$ & $v \sin i$ & $\varepsilon$(He) &
      $\varepsilon$(C) & $\varepsilon$(N) & $\varepsilon$(O) & $\varepsilon$(Mg) & $\varepsilon$(Si) & $\varepsilon$(Fe) \\
       & {\small [K]} & & {\small [km/s]} & {\small [km/s]} & & & & & & & {\small LTE} \\
\hline
$\odot$ & & & & & {\em 10.99} & {\em 8.52} & {\em 7.92} & {\em 8.83} & {\em 7.58} & {\em 7.55} & {\em 7.50} \\
\hline
NGC\,2004/D15 & 22\,500 & 3.80 & 0 & 45 & 10.95 & 8.04 & 6.95 & 8.29 &
      7.35 & 7.04 & 7.30 \\
$\qquad \ \pm$ & {\em 1000} & {\em 0.20} & {\em 2} & {\em 10} & {\em 0.2}
      & {\em 0.2} & {\em 0.2} & {\em 0.2} & {\em 0.2} & {\em 0.2} & {\em 0.3}\\
NGC\,2004 /C16 & 24\,600 & 3.95 & 4 & 60 & 10.94 & 8.10 & 7.05 & 8.43 &
      7.29 & 7.10 & 7.30 \\
$\qquad \ \pm$ & {\em 1000} & {\em 0.20} & {\em 2} & {\em 10} & {\em 0.2}
      & {\em 0.2} & {\em 0.2} & {\em 0.2} & {\em 0.2} & {\em 0.2} & {\em 0.3}\\
NGC\,2004/D3 & 23\,900 & 4.15 & 1 & 70 & 11.00 & 8.15 & 7.05 & 8.40 &
      7.43 & 7.20 & 7.35 \\
$\qquad \ \pm$ & {\em 1000} & {\em 0.20} & {\em 2} & {\em 10} & {\em 0.2}
      & {\em 0.2} & {\em 0.2} & {\em 0.2} & {\em 0.2} & {\em 0.2}  & {\em 0.3}\\
NGC\,2004/C9 & 24\,000 & 4.35 & 0 & 60 & 10.93 & 7.95 & 7.00 & 8.34 &
      7.40 & 7.05 & 7.35 \\
$\qquad \ \pm$ & {\em 1000} & {\em 0.20} & {\em 2} & {\em 10} & {\em 0.2}
      & {\em 0.2} & {\em 0.2}  & {\em 0.2} & {\em 0.2} & {\em 0.2} & {\em 0.3}\\
\hline
$\bar{\varepsilon}{\rm (X)}_{\rm 2004}$ & & & & & 10.96 & 8.06 & 7.01 &
      8.37 & 7.37 & 7.10 & 7.33 \\
$\qquad \ \pm \,\sqrt{\sum\frac{(x_i - \bar{x})^2}{n-1}}$& & & & & {\em 0.03} & {\em 0.09} & {\em0.05} & {\em 0.06} & {\em 0.06} & {\em 0.07} & {\em 0.03}\\
H\,{\sc ii}$_{\rm \,LMC}$ & & & & & 10.95 & 7.90 & 6.90 & 8.40 & --- & 6.70 & ---\\
$\qquad \ \pm$ &  &  &  &  & {\em 0.2} & {\em 0.2} & {\em 0.2}  & {\em 0.2} & --- & {\em 0.2} & --- \\
\hline
NGC\,1818/D1 & 24\,700 & 4.00 & 0 & 30 & 11.04 & 7.80 & 7.40 & 8.46 & 
      7.35 & 7.10 & 7.37 \\
$\qquad \ \pm$ & {\em 1000} & {\em 0.20} & {\em 2} & {\em 10} & {\em 0.2}
      & {\em 0.2} & {\em 0.2} & {\em 0.2} & {\em 0.2} & {\em 0.2} & {\em 0.3} \\
\hline
   \end{tabular} }
\end{table*}

   \begin{figure}
\vspace*{1mm}
   \centering
   \includegraphics[angle=-90,width=0.45\textwidth,clip=]{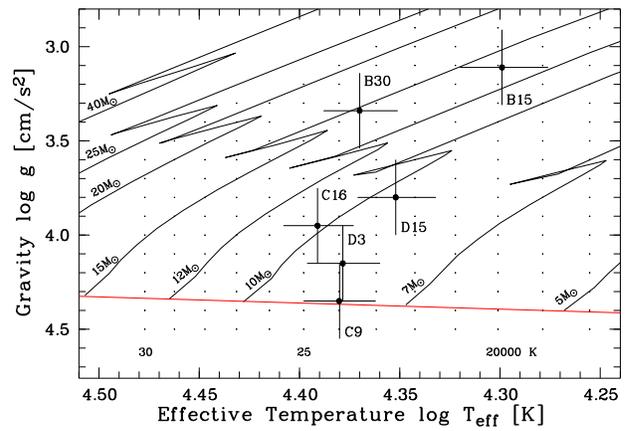}
   \caption{The loci of the programme stars with respect to evolutionary tracks of non-rotating stars for Z\,=\,0.008 (Schaerer et al. \protect\cite{schaerer}). Also shown are the two luminosity class {\sc iii} objects NGC\,2004/B30 and B15 from Paper I. No evolution of chemical elements is detected in comparing the MS stars, whereas B30 from Paper I is clearly enriched in nitrogen and depleted in carbon.}
              \label{loci}%
    \end{figure}

   \begin{figure*}[!ht]
\vspace*{3mm}
   \centering
   \includegraphics[angle=0,width=0.86\textwidth,clip=]{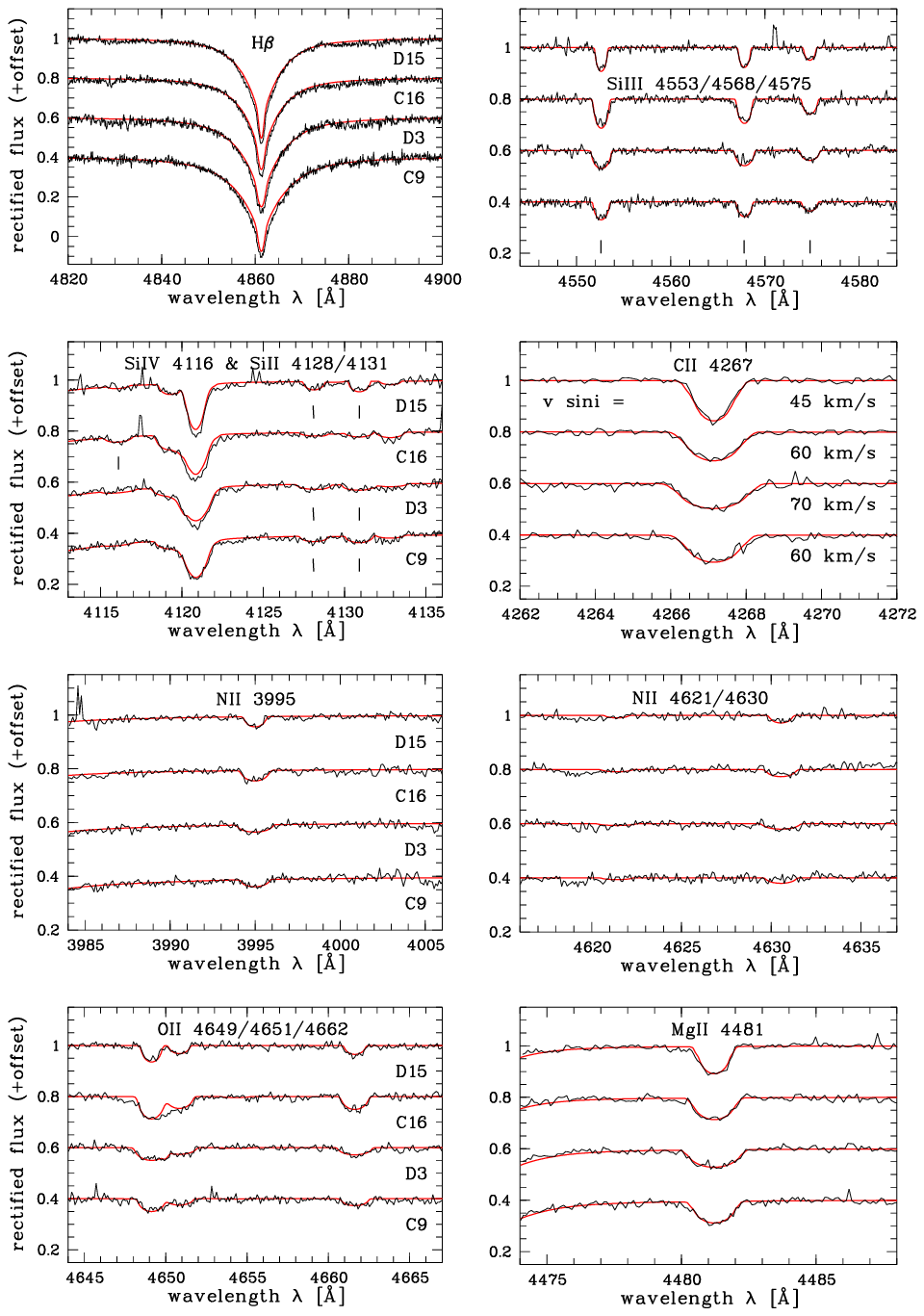}
   \caption{Profile fits for H (gravity determination), Si\,{\sc ii/iii/iv} (temperature and microturbulence indicator), C\,{\sc ii}, N\,{\sc ii}, O\,{\sc ii} (micro\-turbulence indicator) and Mg\,{\sc ii}. The observed noise level can fully account for the difference between theory and observation. Note that the He abundance is not based on He\,{\sc i} 4121\,\AA, since the observed profile seems to contain unidentified blend contributions other that from helium and oxygen} 
              \label{fits}%
    \end{figure*}

\section{Results}
Table \ref{results} gives the stellar parameters and abundances derived for the four programme stars in comparison with H\,{\sc ii}-region data from Garnett (\cite{garnett}). Consistency among the abundances is generally very high with the rms scatter confined to well below 0.1\,dex. Also included is the main sequence star D1 in the LMC cluster NGC\,1818 already analysed in Paper I. For the latter star, the abundances of He, C and N are redetermined on the basis of all available lines and utilizing the new model atoms. Where available, the abundances are compared with measurements from LMC H\,{\sc ii} regions as carried out/compiled by Garnett (\cite{garnett}). Finally, we also include the Solar System composition (Grevesse \& Sauval \cite{grevesse}) as a reference. Note that a recent redetermination of the solar oxygen abundance from [O\,{\sc i}] 6300 based on 3D hydrodynamical computations (Allende Prieto et al. {\cite{allende}) yields $\varepsilon$(O)\,=\,8.69\,$\pm$\,0.05. Holweger (\cite{holweger}) comfirms this trend towards a lower value by deriving $\varepsilon$(O)\,=\,8.74\,$\pm$\,0.08 from eight optical and IR lines taking into account effects due to NLTE and granulation. Thus the difference between oxygen abundances from stars in Orion and the Sun reduces to 0.1\,dex at most (cf. Gummersbach et al. \cite{gummersbach}), but important consequences can also be expected for the evaluation of the [$\alpha$/Fe] ratios in the MCs (see below).\\
Figure \ref{loci} shows the positions of the programme stars in a log\,$T_{\rm eff}$\,--\,log\,$g$ parameter space, sometimes referred to as Kiel diagram. Note that the four MS stars nicely span the full width of the main sequence in the B-star temperature regime. This fact allows us to potentially address chemical evolution of certain elements in these stars during their MS lifetime: no such evolution is, however, found (cf. Table~\ref{results} in which the stars are ordered according to their gravity).\\
Figure \ref{fits} gives examples of profile fits for the most important species of this study: H, CNO, Si and Mg. NLTE effects in iron are currently under investigation. Thus, for the time being, $\varepsilon$(Fe) based on the few Fe\,{\sc iii} lines (W$_\lambda$\,$<$\,40\,m\AA ) visible between 4000\,\AA \ and 5000\,\AA\ is given assuming LTE.

\subsection{Pristine CNO abundances in the LMC}
In comparing the stellar CNO abundances with the nebular ones, one immediately notices several features: the oxygen abundances are in excellent agreement, while both nitrogen and carbon are somewhat more abundant in the stars. As carbon is one of the principle dust-forming elements, interstellar dust depletion is the first reason that comes to mind. Though not of high statistical significance, this offset of 0.16 dex could be interpreted as a manifestation of the fraction of carbon locked up in interstellar dust. Unfortunately, the error limits on both sides prevent us from drawing firm conclusions on this matter. Another observation, however, may lend support to this scenario: a very similar offset is found between the gas-phase carbon abundance of Orion ($\simeq$\,8.4, cf. Garnett et al. \cite{garnettetal}) and that of the Sun (between 8.52 (Grevesse \& Sauval \cite{grevesse}) and 8.59 according to Holweger \cite{holweger}). More work is clearly needed here.\\
Conversely, nitrogen is believed not to deplete into the dust phase in significant amounts (Mathis \cite{mathis}). An enhanced nitrogen abundance in B stars could then be taken as an indication of mixing (internal or external), raising the original (nebular) abundance from which the stars formed some million years ago. The high internal consistency among the $\varepsilon$(N) determinations in the programme stars speaks against this scenario as a likely explanation. Besides, the residual offset with respect to the nebular abundance is hardly of high significance considering the accuracy of 0.2 dex for each of the data sets. We therefore take this result as strong evidence in favour of the extraordinarily {\bf low present-day nitrogen abundance of} \boldmath$\varepsilon$\unboldmath{\bf (N)}\,\boldmath$\simeq$\unboldmath\,{\bf 7.0}. On the square-bracket scale\footnote{[X/Y]$_\star$\,:=\,log($\frac{n_{\rm X}}{n_{\rm Y}})_\star-\log(\frac{n_{\rm X}}{n_{\rm Y}})_\odot$} this corresponds to [N/H]\,$\le$\,$-$0.92 at a metallicity of [Fe/H]\,$\simeq$\,$-$0.2 or [N/Fe]\,$\simeq$\,$-0.7$. [N/Fe]\,$\simeq$\,0 is only reached {\em after} the first dredge-up on the red-giant branch, as already emphasized by Venn ({\cite{venn1}).\\
Finally, we comment on oxygen-related matters: there can be little doubt that the {\bf present-day LMC oxygen
abundance} is {\bf close to 8.40}. With a mean nitrogen abundance of 7.01, our
mean oxygen value of 8.37 results in log(N/O)\,=\,$-$1.36. Therefore,
even though the metallicity of the LMC is not so different from
solar, its N/O ratio is among the lowest found in the local universe, 
as derived from Galactic and extragalactic
nebular and stellar abundances (cf. Henry et al. \cite{henry}).
Indeed, within the chemical evolution model favoured by Henry et al.,
which attributes nitrogen production entirely to low and intermedate-mass stars, this implies that the present-day nitrogen abundance of the LMC is still
dominated by primarily produced nitrogen, which is consistent with
the metallicity-dependent nitrogen yields of van den Hoek \&
Groenewegen (\cite{hoek}).
   \begin{figure*}
\vspace*{2mm}
   \centering
   \includegraphics[angle=0,width=0.97\textwidth,clip=]{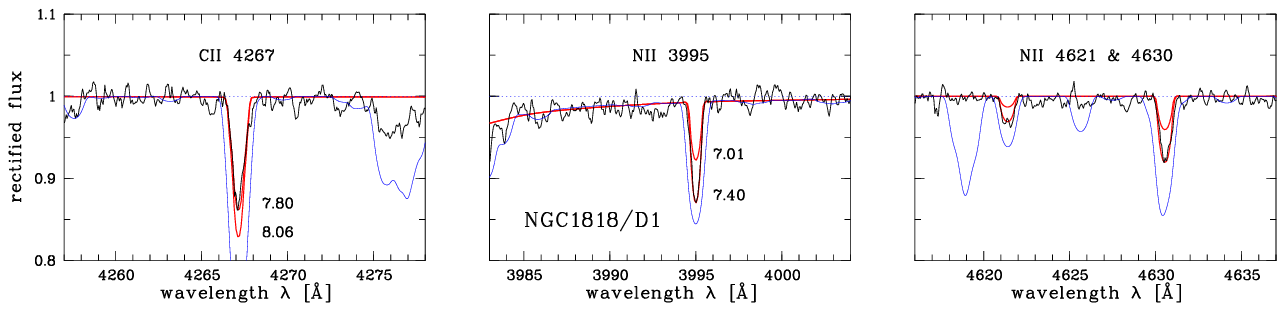}
   \caption{Carbon and nitrogen profile fits for NGC\,1818/D1: the best-fit value is compared to the mean value as derived from the four MS stars in NGC\,2004. The thin line represents the LTE utility {\sc synspec} ($T_{\rm eff}=25\,000\,{\rm K}, \log\,g=4.0$, $\xi$\,= 10\,km/s, solar composition, cf.\ http://feros.lsw.uni-heidelberg.de/cgi-bin/websynspec.cgi) which is used to locate the continuum. Note that some of the line strengths are greatly overestimated under LTE conditions (e.g. C\,{\sc ii} 4616/4619 and 4625).}
              \label{D1}%
    \end{figure*}

\subsection{The $\alpha$-elements}\label{alpha}
Due to a state-of-the-art redetermination of the solar oxygen abundance (see above) it has become unclear what [O/Fe] ratio $\varepsilon$(O)\,=\,8.40 corresponds to. In fact, this question concerns not only oxygen, but all other $\alpha$-elements (Ne, Mg, Si, S, Ca and possibly Ti) in the sense that oxygen is often taken to be indicative of the behaviour of these elements. Thus the whole question of whether or not the [$\alpha$/Fe] ratios in the MCs are vastly different from those found in the Galactic thin disk at the appropriate metallicities must be re-evaluated in the light of these recent developments: From the three $\alpha$-elements we have access to, we derive [$\alpha$/Fe]\,$\simeq$\,$-0.15$ using the new solar oxygen abundance. Even if we account for a lower NLTE Fe abundance, [$\alpha$/Fe] is still significantly lower than in the Galactic thin disk: $+0.10$ at [Fe/H]\,$\simeq$\,$-0.3$ (Fuhrmann \cite{fuhrmann1}, \cite{fuhrmann2}).\\
For silicon we find a relatively clear offset of stellar silicon abundances relative to that from H\,{\sc ii} regions of 0.4\,dex which could be significant. The LMC giants of Paper~I show a similar offset ($\simeq$\,0.3\,dex) which can most naturally be explained as the fraction of silicon locked up in interstellar grains (silicates).} 

\subsection{Iron}
On average, the iron abundances presented in Table \ref{results} are merely 0.17\,dex below the solar value which is quite a high value for LMC objects. We reemphasize, however, that {\em LTE results} are given in Table \ref{results}. First results of the new NLTE computations indicate that the LTE results constitute {\em upper} limits to the NLTE abundances. Thus, the high iron abundances is most likely caused by the inappropriateness of LTE for stars as hot as 25\,000\,K. For the time being, we account for this fact by assigning larger error bars ($\pm$\,0.3\,dex).   

\subsection{Isochrone ages}
In comparison with evolutionary tracks, the stellar para\-meters may serve to constrain the cluster age which has only been determined from colour-magnitude diagrams (CMD) so far. Biases are certainly present in both cases: MS fitting in CMDs will be prone to Be stars and rapid rotators ``above'' the MS mimicking a hotter turnoff and artificially lowering the derived age (cf. Grebel et al.\ 1996). Additionally, (differential) reddening is always an issue. In the case of spectroscopic ages derived from individual stars, unknown cluster membership might introduce a bias, not even the direction of which is known. Advantageously, reddening does not enter.\\
Having shown that our programme stars are bona-fide unmixed objects, we are hopeful that the derived ages are at least not affected by rotation. As can be seen from Fig.~\ref{loci}, the ages scatter substantially, e.g. C9 is fully compatible with age zero. It is unclear whether or not this hints at multiple star formation episodes for NGC\,2004, as proposed by Caloi \& Cassatella (\cite{caloi}) based on IUE data. Our best estimate comes from the most evolved MS target D15 close to the turnoff where the age resolution is highest: $19\,$Myr $\leq\,t_{\rm NGC\,2004}\,\leq\,24\,$Myr. A CMD analysis by Bencivenni et al. (\cite{bencivenni}) points towards a much lower age of 8\,Myr confirming an earlier result by Hodge (\cite{hodge}). Note, though, that these estimates are based on $B-V$ colours which are a rather insensitive indicator of turnoff temperatures and prone to contamination with Be stars. More recently, Keller et al. (\cite{kelleretal}) utilized the HST filters F160BW (far UV), F555W (``$v$'') and F656N (H$\alpha$), separated the Be population and derived an age for the cluster of 16\,$\pm$\,4\,Myr, in much better agreement with our result.     
\subsection{A fresh look at NGC\,1818/D1}
Having redetermined the He, C and N abundance of this star, it is worthwhile taking another look at it: even though both $\varepsilon$(He) and $\varepsilon$(N) have decreased due to better statistics and more sophisticated modelling, the signature indicative of mixing with CN-cycled material remains intact. More importantly, we now have pristine B star abundances to compare it to (see Fig.~\ref{D1}).\\ 
The agreement in the O, Mg, Si and Fe abundance with NGC\,2004/C16 (the object closest to D1 in terms of stellar parameters) is remarkable, yet both carbon and nitrogen are offset by $\geq$\,0.3 dex in opposite directions. We refrain from rephrasing the discussion already presented in Paper I, but simply stress that NGC\,1818/D1 remains a prime candidate for a MS B stars that has undergone rotational mixing. We urge observations of boron to be performed for this star which according to Fliegner et al.\ (\cite{fliegner}) have the potential of distinguishing between external contamination (e.g.\ binary interaction) and an internal process like rotational mixing.  

\section{Conclusions}
We have presented detailed chemical abundances for four MS B stars in NGC\,2004. This study doubles the number of MS B stars in the LMC for which accurate abundances have been derived by means of high-resolution spectroscopy. None of the stars shows any indication of effects due to binary interaction or rotational mixing allowing us to derive what we consider to be the first data set of {\em unaltered present-day CNO abundances} from LMC stars.
\begin{itemize}
\item
We confirm the extraordinarily low LMC nitrogen abundance previously
found from H\,{\sc ii}-region studies to within 0.1 dex. This value is more than 
0.5\,dex below average values found in the Galactic thin disk at LMC metallicities (Liang et al. \cite{liang}, cf. their Fig.~10). It implies an enrichment history for the two environments different from one another and -- within the chemical evolution model of Henry et al. (\cite{henry})~-- a dominance\linebreak of primary nitrogen production in the LMC {\em until\linebreak today}.
\item
In our programme stars, carbon is 0.16\,dex above the nebular value. While this is the direction in which dust depletion would act, the offset found is hardly significant considering the error limits on both sides. More work is needed to reduce both random and systematic errors to draw definitive conclusions on dust depletion fractions.
\item
The B-star oxygen abundance (also from B giants, cf. Paper I) is in excellent agreement with the nebular value. In the absence of systematic offsets between the two data sets, little nebular oxygen seems to be tied up in grains at LMC metallicities.
\item
The stellar silicon abundance is offset from the nebular one by 0.4\,dex. 
This might imply that a major fraction of the interstellar silicon in the LMC is bound in grains.
\end{itemize}
This work illustrates the potential of abundance analyses of hot stars based on high-quality spectra and sophisticated input physics. With modern multi-object spectrographs (e.g. GIRAFFE on VLT UT2 going on-line in 2002) we will be able to analyse some 100 slowly rotating B stars in each Galactic and MC cluster like NGC\,2004. This will not only result in reduced {\em random errors} (which we will be able to address truly statistically for the first time), but also prove or refute the importance of rotational mixing for this class of stellar objects. In particular, a sequence consisting of Galactic, LMC and SMC B stars could  settle the r\^{o}le metallicity, rotation and mass loss play in hot-star evolution. As for the H\,{\sc ii}-region data, more objects will have to be observed to beat down the errors on that side as well.\\
{\em Systematic errors} --~arguably more important than random ones at the current level of accuracy~-- will have to be addressed by an integral approach: by studying the stellar and gas component of H\,{\sc ii} regions in tandem. Both fields of research would undoubtedly profit from such an endeavour.

\begin{acknowledgements}
      AJK wishes to thank I.\,Appenzeller for making this work possible financially: the hospitality of the LSW during April and May 2001 was well-appreciated.\\
We thank an anonymous referee for valuable comments on the initially submitted version.\\
SCK's work at LLNL was performed under the auspices of the U.S. Department of Energy by the University of California Lawrence Livermore National Laboratory under contract No. W7405-ENG-48.
\end{acknowledgements}

\end{document}